\documentclass[a4paper,twoside]{article}
\newcommand{\affil}[1]{$^{\rm #1}$}
\usepackage[authoryear]{natbib}
\usepackage{aas_macros}
\bibpunct{(}{)}{;}{a}{}{,}
\usepackage{graphicx}
\usepackage{amsmath}
\usepackage{aas_macros}
\usepackage{url}
\usepackage{float}
\date{} 
\title{\large\bf\flushleft 2D-1D Wavelet Reconstruction As A Tool For Source Finding In Spectroscopic Imaging Surveys}
\author{\parbox{\textwidth}{\flushleft
\vspace{-0.5cm}
{\it Lars Fl\"oer \affil{A,C} and Benjamin Winkel \affil{B}}\\
\vspace{0.4cm}
{\small \affil{A}\,Argelander-Institut f\"ur Astronomie, Auf dem H\"ugel 71, 53121 Bonn, Germany}\\
{\small \affil{B}\,Max-Planck-Institut f\"ur Radioastronomie, Auf dem H\"ugel 69, 53121 Bonn, Germany}\\
{\small \affil{C}\,Email: lfloeer@astro.uni-bonn.de}}}
\begin{document}
\twocolumn[
\begin{changemargin}{.8cm}{.5cm}
\begin{minipage}{.9\textwidth}
\vspace{-1cm}
\maketitle
%
%
\small{\bf Abstract:} Today, image denoising by thresholding of wavelet coefficients is a commonly used tool for 2D image enhancement. Since the data product of spectroscopic imaging surveys has two spatial and one spectral dimension, the techniques for denoising have to be adapted to this change in dimensionality. In this paper we will review the basic method of denoising data by thresholding wavelet coefficients and implement a 2D-1D wavelet decomposition to obtain an efficient way of denoising spectroscopic data cubes. We conduct different simulations to evaluate the usefulness of the algorithm as part of a source finding pipeline.

\medskip{\bf Keywords:} methods: data analysis --- techniques: image processing --- techniques: spectroscopic

\medskip
\medskip
\end{minipage}
\end{changemargin}
]
\small

\section{Introduction}

\begin{table}[t]
\begin{center}
\caption{Number of abstracts on ADS containing the word ``wavelet'' for given date ranges.}
\begin{tabular}{cc}
\hline
Years & Number of Abstracts \\
\hline
Until 1995 & 251 \\
1996 - 2000 & 679 \\
2001 - 2005 & 1221 \\
Since 2006 & 1797 \\
\hline
\end{tabular}
\end{center}
\label{tab:adsresults}
\end{table}%

The usage of the wavelet transform in astrophysics has become very popular in recent years. Table \ref{tab:adsresults} compiles the number of publications on ADS\footnote{NASA Astrophysics Data System, \protect\url{http://adswww.harvard.edu/}} in a given range of years that have the word ``wavelet'' contained in their abstract. Clearly, the usage of wavelets has gained popularity fast. Typical applications for wavelet-transform-based methods are morphological separation of sources in images and noise removal. The success of wavelet based methods in astrophysics is in part due to the fact, that astrophysical data often contains information on different angular or spectral scales. For example, an optical image of a galaxy contains compact, bright stars as well as extended and faint emission from the bulge and spiral arms. Multi-scale methods, such as the wavelet transform, allow to investigate the different scales of an image separately \citep{5299269}.

The most widely used type of wavelet transformation is the so called undecimated or redundant, isotropic wavelet transformation. This is in part due to the algorithmic simplicity of the method but also because undecimated wavelet transforms have proven to be more efficient for noise removal then their decimated counterpart. Apart from that, they also provide a number of computational advantages when reconstructing an image from a subset of its wavelet coefficients \citep{Starck:2010uf}.

In this paper we review the basics of denoising based on the undecimated wavelet transformations in Section \ref{sec:wtdenoise} and present an extension of the wavelet transform to three dimensional data as proposed in \cite{2009A&A...504..641S}. Section \ref{sec:impl} describes the implementation of the transform in C++ along with a description of where the implementation departs from the original algorithm. A first application of the algorithm is shown in Section \ref{sec:application}, where we use the algorithm to implement a source finder and test the performance on simulated H\,{\sc i} galaxies. We close the paper with a summary and an outlook on future applications and potential improvements to the algorithm.

\section{Wavelet denoising}\label{sec:wtdenoise}

The isotropic, undecimated wavelet transform (IUWT) decomposes data $D(x)$ into $J+1$ subbands
\begin{equation}
\label{eq:wdec}
 D(x) = c_{J}(x) + \sum_{j=1}^{J} w_{j}(x)
\end{equation}
where $c_{J}$ is a smooth version of the data and the details at position $x$ and scale $j$ are contained in the wavelet coefficients $w_{j}(x)$. The IUWT can be efficiently calculated by using the so called ``algorithme \`a trous'' \citep{1989wtfm.conf..286H}. To calculate the IUWT one needs to convolve the input data with increasingly larger kernels. To calculate the next convolution, the algorithme \`a trous convolves the previously convolved data again with the same kernel with $2^j$ zeros inserted between the kernel values. For multidimensional transforms this insertion of zeros is done isotropically in all dimensions. This allows efficient calculation of even the largest scales.

At each step, two consecutive convolved versions of the data $c_j(x)$ and $c_{j+1}(x)$ are used to calculate the wavelet coefficients $w_{j}(x) = c_{j}(x) - c_{j+1}(x)$. The number of scales is usually chosen to be $\lfloor\log_{2}N_{i}\rfloor$, where $N_{i}$ is the number of samples per dimension in a data set, e.g. an image with $N_{1}\times N_{2}$ pixels. A more detailed description can be found in \cite{Starck:2010uf}.

When using wavelets to denoise data, one assumes, that the signal in the data, e.g. the sources, can be described by only a few relevant coefficients in each of the detail subbands $w_{j}(x)$, i.e. that the signal is sparse in a given wavelet representation. Consequently, one can try to detect only the relevant coefficients and reconstruct the image from those.

The detection is usually based on estimating the standard deviation $\sigma_{j}$ of the coefficients in a given subband and only take the coefficients with absolute values above a certain threshold $t\sigma_{j}$ to be significant. $t$ is usually chosen to be between $3$ and $5$. Then, if one applies Equation \ref{eq:wdec}, with all insignificant coefficients set to zero, one obtains a denoised approximation of the data.

Since this nonlinear denoising benefits greatly if iterated a few times, \cite{1995A&AS..112..179M} developed the notion of a multi-resolution support $M$, that contains information about whether the data has a significant coefficient at a given location and scale. The multi-resolution support is defined as follows:
\begin{equation}
 M(x,j) \left\{
 \begin{array}{ll}
 1 & \text{if $w_{j}(x)$ is significant}\\
 0 & \text{else}
  \end{array}\right.
\end{equation}

Using this multi-resolution support, one can implement the following iterative reconstruction scheme:
\begin{enumerate}
\item Detect all significant coefficients $w_{j}(x)$ and store this information in the multi-resolution support $M(x,j)$.
\item Calculate the IUWT of the data $D$ and reconstruct the image only from the coefficients that belong to $M$ to obtain $\tilde D$.
\item Calculate the residual $R = D - \tilde D$.
\item Calculate the IUWT of $R$ and again only retain the coefficients that belong to $M$. Add this reconstruction to $\tilde D$.
\item Go to step 3 until the desired number of iterations is reached.
\end{enumerate}
In practice a small number of iterations ($<$\,10) is sufficient. Many examples of how iteration improves the denoising process can be found in \cite{Starck:2010uf}.

\subsection{Extension to 2D-1D data}

The aforementioned decomposition and reconstruction works very well if the relevant signal in the data is isotropic or nearly isotropic. This is true for most 1D and 2D astrophysical data like spectra and images. In the case of imaging spectroscopic surveys like the past ``H{\sc i} Parkes All-Sky Survey''(HIPASS; \citealt{2001MNRAS.322..486B,2004AJ....128...16K}), ongoing ``Effelsberg-Bonn H{\sc i} Survey'' (EBHIS; \citealt{2011AN....332..637K,2010ApJS..188..488W}), the ``Arecibo Legacy Fast ALFA Survey'' (ALFALFA; \citealt{2005AJ....130.2598G}) and future surveys with WSRT+Apertif \citep{2009wska.confE..70O} and the Australian SKA Pathfinder (ASKAP; \citealt{2008ExA....22..151J}), the data is three dimensional with two angular and one spectral dimension, which is referred to as a data cube.
Since spatially unresolved sources can still be resolved spectrally, sources generally do not have the same size among the three different axes of the data cube. This leads to an anisotropy, which makes isotropic denoising schemes inefficient, since the wavelet decomposition does not match the natural shape of the sources very well. Nonetheless, the sources can be considered partly isotropic in each individual spectral slice (channel map) and are also approximately isotropic along each line of sight. It is therefore beneficial to split the wavelet decomposition up into a two dimensional angular and a one dimensional spectral part.
 
The theoretical foundation for this 2D-1D transformation is laid out by \cite{2009A&A...504..641S}, which apply a 2D-1D denoising to data from the Fermi LAT \citep{2009ApJ...697.1071A}. Fermi LAT data has either two angular and one spectral or two angular and one temporal domain. Even though very different from radio astronomical observations in its noise characteristics, it is similar to data from imaging spectroscopic surveys in terms of the dimensionality. 

To calculate a wavelet representation that accounts for this difference in axis type, they first calculate a 2D IUWT of each channel map and subsequently apply a 1D IUWT along each pixel of this wavelet coefficient data cube. When applying this decomposition with $J_{1}$ angular and $J_{2}$ spectral scales one arrives at a decomposition of the form
\begin{align}\label{eq:2d1dwdec}
 D(x) &= c_{J_{1},J_{2}}(x)\nonumber\\ 
 &+ \sum_{j_{1}} w_{j_{1},J_{2}}(x) + \sum_{j_{2}} w_{J_{1},j_{2}}(x) \nonumber\\
 &+ \sum_{j_{1},j_{2}} w_{j_{1},j_{2}}(x)\quad.
\end{align}
Analogous to Equation \ref{eq:wdec}, $c_{J_{1},J_{2}}(x)$ is the smooth version at angular scale $J_{1}$ and spectral scale $J_{2}$. The second row contains the coefficients that arise from either spectral decomposition of the smooth angular scale or angular decomposition of the smooth spectral scale. The last sum of coefficients $w_{j_{1},j_{2}}$ contains the detail of the data at angular scale $j_{1}$ and spectral scale $j_{2}$.

To implement an iterative denoising scheme as described above, one again has to construct a five dimensional (two angular dimensions, one spectral dimensions, and two scale indices), multi-resolution support $M(x,j_{1},j_{2})$. Once all significant coefficients are detected, the iterative reconstruction can be applied as described in the previous section.

\section{Implementation}\label{sec:impl}

\subsection{Scale selection}\label{sec:scaleselect}

In general, the denoising of data is performed by decomposing the image with the maximum amount of scales, i.e. $J_{i} = \lfloor\log_{2}N_{i}\rfloor$. There are however certain advantages in only using a subset of decomposition scales for both the angular and spectral regime. This is especially true when the denoised image only serves as a mask to find sources in the original data and complete flux reconstruction is not of importance.

The signature of a galaxy in neutral hydrogen surveys is typically small compared to the dimensions of the data cube. It is therefore unlikely that one will miss any sources when leaving out all decomposition scales that belong to larger spectral or angular scales. Especially for single dish observations, the information contained at the very large scales is most likely due to baseline errors or radio frequency interference (RFI). By neglecting those larger scales, one can suppress those errors in the reconstruction and extract sources even from suboptimal data. This property is investigated in Section \ref{sec:robustness}.

For this reason, the reconstruction in our algorithm is done with a ``physical'' subset of angular and spectral scales that are likely to contain the signal of sources. Likewise, the coefficients $w_{J_{1},j_{2}}$, $w_{j_{1},J_{2}}$ and the smooth data $c_{J_{1},J_{2}}$, which all contain the information at the largest scales, are not taken into account, and are not part of the reconstructed data.

What scales are to be considered physical depends on the type of source one is looking for. In this paper we mainly focus on extragalactic objects, namely H{\sc i} galaxies, which are typically not larger than a few arc minutes. What scales then contain the desired objects depends on their typical angular and spectral size and the respective sampling of the dimensions on the voxels of the data cube. 

Additionally, we only reconstruct the data from the positive wavelet coefficients. This approach is different from the usual iterative approach, where all significant wavelet coefficients are used and the negative values of the reconstruction are set to zero at each iteration. We noticed, that by choosing only the positive coefficients and using mathematical morphology (see next section) we suppress the artifacts that arise during partial reconstruction from wavelet coefficients \citep{4060954}. If unsuppressed, these artifacts make the usage of the reconstruction as a mask for source finding difficult, as they can also lead to merging of sources. On the other hand, this positivity constraint makes searching for negative features, e.g. absorption lines, impossible.

\subsection{Mathematical morphology}

Another advantage of storing the information of the significant coefficients in the multi-resolution support is, that one can perform mathematical morphology on it \citep{Serra:1982fk}. Generally, data cubes are created in a way, that the sampling of the telescope beam fulfills the Nyquist sampling theorem \citep{nyquist}, meaning that it is sampled on at least two pixels in every direction (including the diagonal). This means that real sources are larger than a single pixel in the angular dimension and are most likely also sampled in more than one spectral channel. Furthermore,  it is well known that significant structures propagate through the different scales of the IUWT. Sources will therefore be present in multiple adjacent coefficients in the three dimensions of the data as well as adjacent spatial and spectral scales and form connected regions in the five-dimensional multi-resolution support.

To further suppress the noise in the reconstruction we perform a five-dimensional morphological opening of the multi-resolution support. Morphological opening consists of the successive application of an erosion followed by a dilation. The former removes elements from the multi-resolution support if one of its neighbors (in all five dimensions) is 0 and the latter does the opposite, i.e. adding elements to the multi-resolution support if one of its direct neighbors is 1. This amounts to a successive shrinking and growing of objects in the five-dimensional multi-resolution support.

Objects spanning all five dimensions of the multi-resolution support are not affected by this operation. However, objects that do not span all five dimensions, and are therefore likely to be noise artifacts, are removed. This allows us to use much lower thresholds of typically $1.5\sigma_{j}$ during reconstruction. For the purposes of source finding this is both an increase in sensitivity as well as reliability.

\subsection{Memory layout and processing}

The described 2D-1D denoising has been implemented in C++ using the \`a trous algorithm in both the two dimensional angular, as well as the one dimensional spectral transformation. The complete storage of all wavelet coefficients would take $J_{1}\times J_{2}$ the amount of memory of the original data, which can easily exceed the available computing resources for the typical size of a data cube of several hundreds of MByte. Here, we deal with this major issue by performing the reconstruction on-the-fly.  Such a serialized method only needs to store the angular smoothed version $c_j$, the angular wavelet coefficients $w_{j_1}$, and the reconstruction $\tilde D$. This way, the memory consumption of the algorithm is greatly reduced being now independent on the number of angular and spectral scales analyzed.

Another memory concern is the size of the multi-resolution support, that has to store $N_{1}\times N_{2} \times N_{3} \times J_{1} \times J_{1}$ boolean values, where $N_{i}$ is the size of the data cube in pixels along the $i$th axis. For this purpose, the Standard Template Library (STL) for C++ implements a specialized container, that is able to store boolean values as individual bits rather than bytes, which makes the memory footprint of the multi-resolution support acceptable.

The splitting of the different wavelet transformations makes this denoising scheme a prime candidate for parallel computing. Using the OpenMP\footnote{\url{http://openmp.org}} library, the angular wavelet decomposition of each spectral channel as well as the spectral wavelet decomposition of each line-of-sight was implemented to be computed in parallel.

\section{Simulations}\label{sec:application}

To examine the various aspects in the sections below, we created 1000 simple H\,{\sc i} galaxy templates using the GIPSY\footnote{\url{http://www.astro.rug.nl/~gipsy/}} task \texttt{galmod}. The galaxies were simple disks with random inclination and maximum rotational velocity while keeping the overall brightness profile and rotation curve fixed.

Noise was generated according to the specifications of the WALLABY\footnote{\url{http://www.atnf.csiro.au/research/WALLABY/}} survey (\citealt{wallabyprop}; also see Koribalski 2011, this PASA issue). The models were convolved with a Gaussian beam of approximately $30^{\prime\prime}$ and inserted into data cubes with an rms noise of $1.8\rm\,mJy/beam$. The exact specifications are however not important for the simulations since all tested quantities are given in terms of signal-to-noise ratios and the algorithm only operates on the pixel grid of the data. A difference in beam sizes should therefore yield the same results if the beam is sampled on the same number of pixels.

The algorithm was run on multiple data cubes of 300 by 300 pixels and 600 channels size that each contained 20 random galaxies at random positions.

\subsection{Source scaling}

\begin{figure}[t]
\begin{center}
\includegraphics[width=1\columnwidth,trim=10 0 0 0,clip=true]{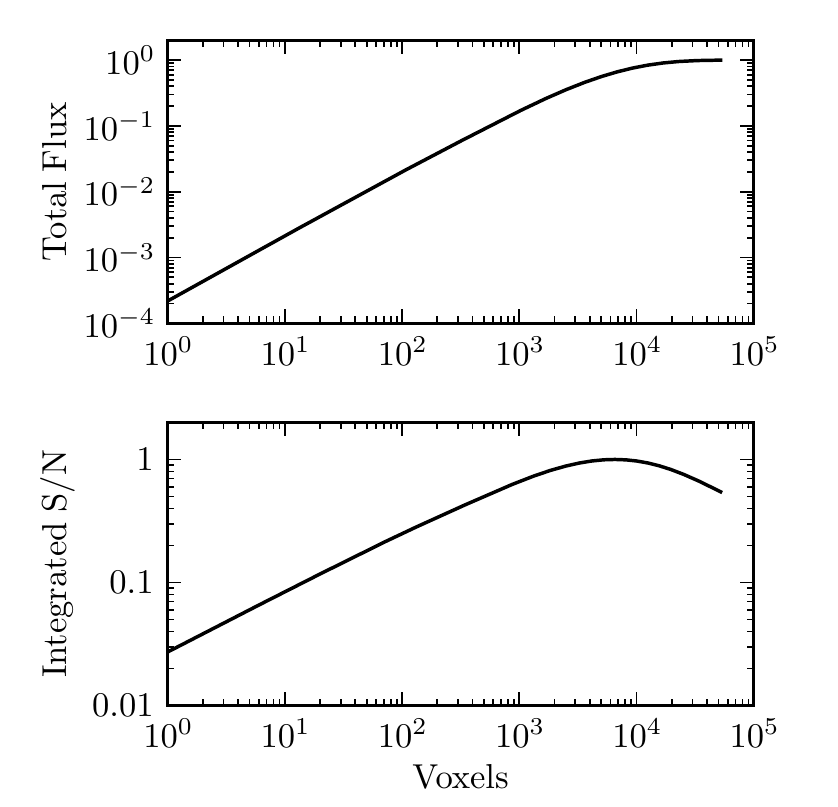}
\caption{Normalized total flux (top) and normalized integrated signal-to-noise ratio (bottom) as a function of the number of voxels added up, ordered by flux in descending order, for a given source model.}\label{fig:isnrplot}
\end{center}
\end{figure}

Since the proposed algorithm is sensitive to the complete source signal in the data as opposed to e.g. the peak flux, we scaled each of the 1000 galaxies to a fixed set of integrated signal-to-noise ratios. Since the integrated signal-to-noise ratio (ISNR) is dependent on the volume over which it is calculated, we first determined the optimal volume for each of our models. This was done by starting with the brightest voxel of the model and successively adding the next fainter one. This way, both the total flux and ISNR will increase as a function of the number of voxels added. At a certain point, the flux in the added voxels becomes very low, since one adds the faint ``outskirts'' of the model. At this point the ISNR will go down since one adds more noise than source flux. This behavior can be seen in Figure \ref{fig:isnrplot}. The flux at this optimal ISNR is then scaled to yield the desired ISNRs.

\subsection{Example reconstruction}

\begin{figure}[t]
\begin{center}
\includegraphics[width=1\columnwidth,trim=15 0 7 0,clip=true]{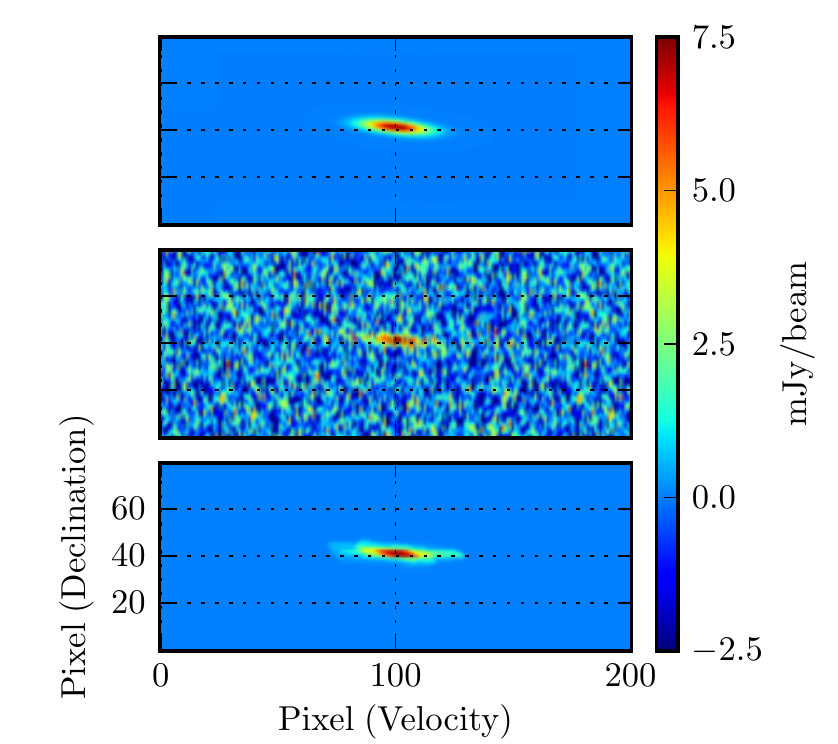}
\caption{Example of a wavelet reconstruction by the described algorithm. From top to bottom the panels show: source model only, source model with added noise, reconstruction by the algorithm.}\label{fig:reconstruction}
\end{center}
\end{figure}

\begin{figure}[t]
\begin{center}
\includegraphics[width=1\columnwidth,trim=13 0 22 0,clip=true]{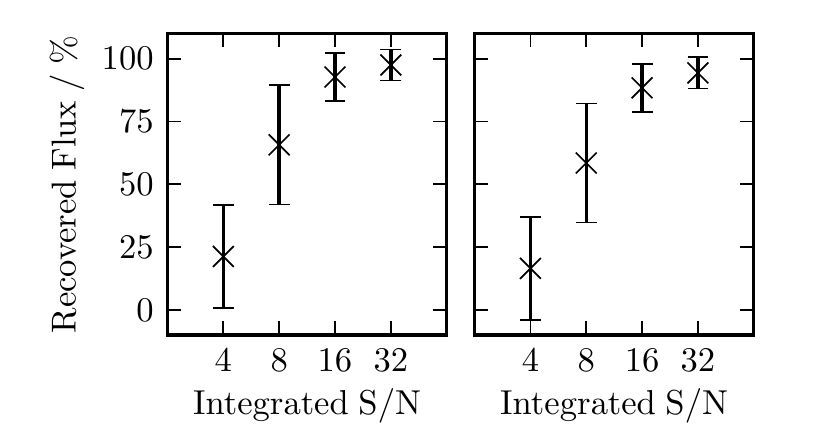}
\caption{Recovered flux as a function of ISNR. The crosses show the mean of 1000 galaxy models and the error bars indicate the standard deviation in each bin. The left panel shows the recovered flux as obtained from the wavelet reconstruction. The right panel shows the recovered flux when applying the wavelet reconstruction as a mask for data and calculating the flux from the masked data.}\label{fig:recdatflux}
\end{center}
\end{figure}

Figure \ref{fig:reconstruction} shows the typical result of a reconstruction by the described algorithm. The top panel shows one of our scaled templates. In the middle panel, the simulated noise was added. The bottom panel shows the same data cube after the application of the described denoising algorithm. In general, the reconstruction does not restore the full flux of the inserted model and also has a changed appearance as compared to the model. This is especially true for low signal-to-noise sources, since the reconstruction becomes limited by noise. For more pronounced sources there is however a good correlation between model flux and reconstructed flux. This is shown Figure \ref{fig:recdatflux}: we plot recovered flux of the reconstruction (left panel) as well as the flux recovered from the data when using the reconstruction as a mask (right panel) as a function of ISNR. Especially for ISNR 32 and 16, the recovered flux matches the model quite well. It is also interesting to note, that the flux from the reconstruction seems to be closer to the true flux than the flux calculated from the data. In any case, this evaluation also shows, that the masks obtained from the reconstruction should not be used as the final masks without further treatment.

\subsection{Robustness}\label{sec:robustness}

\begin{figure}[t]
\begin{center}
\includegraphics[width=1\columnwidth,trim=15 0 7 0,clip=true]{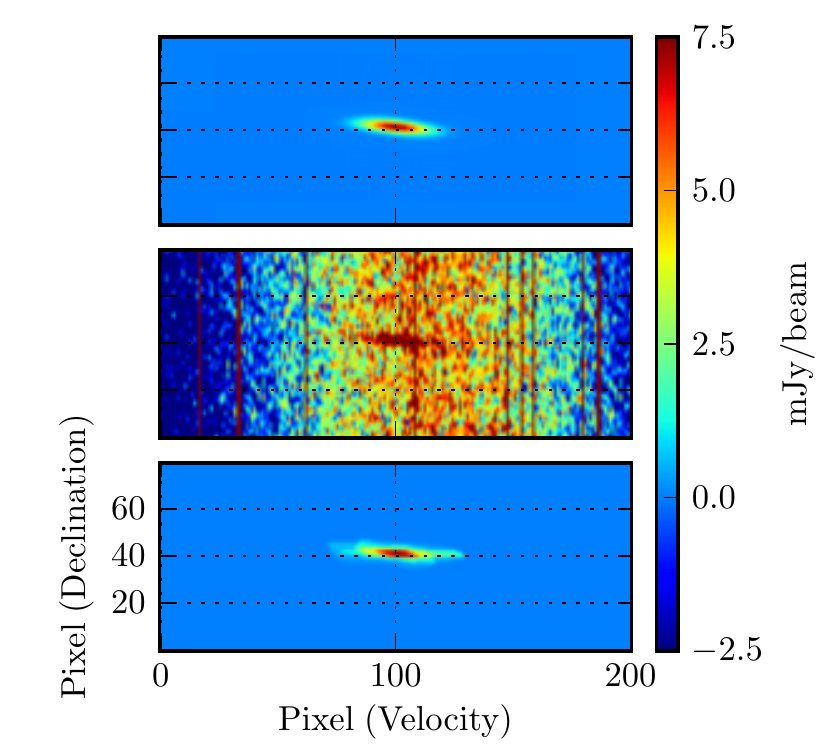}
\caption{Same as Figure \ref{fig:reconstruction} but for corrupted data. In addition to the simulated noise, the middle panel shows a sinusoidal varying baseline and added RFI spikes.}\label{fig:robustness}
\end{center}
\end{figure}

Since real data does not usually contain ideal noise and sources, we evaluated the robustness of the proposed algorithm against two common types of data defects: baseline ripple and RFI. 

To simulate these effects, we added a sine wave to one simulated data cube with a varying phase along one angular axis. To simulate the presence of RFI we inserted 30 single-channel spikes in the data and reran the wavelet reconstruction. The result can be seen in Figure \ref{fig:robustness}. Clearly, the wavelet reconstruction is not affected by the rather severe presence of RFI and baseline ripple. This is because both the baseline ripple and RFI are present in scales different from the scales of the sources. By carefully selecting which scales to reconstruct the data from, we can exclude many of such defects.

\subsection{Completeness and reliability}

The two main measures of goodness of a source finder are its completeness as a function of source signal and the corresponding reliability. The completeness is expressed as the percentage of sources that have been positively identified by the source finder. The reliability is calculated as the number of true sources divided by the total number of objects found by the source finder and gives a measure of the probability that a given object is indeed a source or a false positive.

To test the performance of the algorithm as a source finder, we set up a simple source finding pipeline by using several functions from the SciPy\footnote{\url{http://scipy.org}} package \texttt{ndimage}. After the wavelet reconstruction, the \texttt{ndimage} functions \texttt{label} and \texttt{find\_objects} are used to generate the objects. For this purpose \texttt{label} searches the data cube for connected objects, i.e. regions where the flux is greater than zero, and marks each region with a unique number. \texttt{find\_objects} then generates a list of slices that  each fully contain one of the labeled objects. Those slices are then used to calculate various parameters like the total flux of the reconstructed object $F_{R}$, the total flux in the original data $F_{D}$ when applying the reconstruction as a mask and various shape parameters like the size in channels. To check whether a given object is a true detection, we use a noise free version of the same data set and check for intersections with the noise free sources above 20\% of the peak flux of a galaxy.

\subsubsection{False positives}

\begin{figure}[t]
\begin{center}
\includegraphics[width=1\columnwidth,trim=15 0 15 0,clip=true]{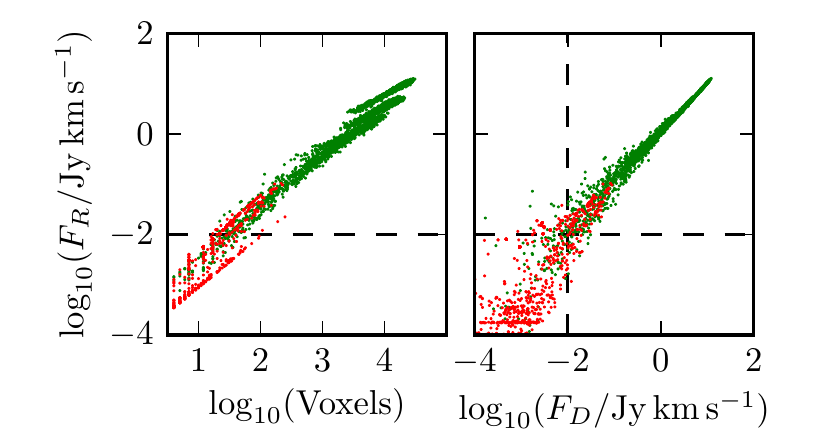}
\caption{Distribution of the detection parameters $F_{R}$, the flux in the reconstruction, $F_{D}$, the flux in the detection as measured on the original data and the number of voxels a given object occupies. The green points indicate true detections, the red points false detections. The dashed lines indicate $10\rm\,mJy\,km\,s^{-1}$ for $F_{R}$ and $F_{D}$, respectively.}\label{fig:distribution}
\end{center}
\end{figure}

To achieve a reasonably high completeness even for very faint sources, one has to use very low thresholds which will lead to an increasing number of false positives. After the contrast enhancement, the identification of false positives is a key task of every source finder.

Since the source of the false positives, i.e. noise peaks, is greatly suppressed by the algorithm, they mostly stem from the larger wavelet coefficients where the noise peaks are spread out over a sufficiently large volume to not be removed by the morphological opening. This leads to a very low reconstructed total flux $F_{R}$ and they are therefore easily separated from the real sources.

Figure \ref{fig:distribution} shows the correlation for three parameters from the simulation with resolved sources (see Section \ref{sec:simres}), all directly measured from either the reconstructed or the original data. It is evident that all false positives cluster in one region of the respective plots and that they exhibit very low $F_{R}$ and $F_{D}$. Therefore, by applying a simple cut in the parameter space of the detections, the number of false positives can be greatly reduced without sacrificing much of the completeness. By only taking sources that have both $F_{R}$ and $F_{D}$ larger than $10\rm\,mJy\,km\,s^{-1}$, we exclude 96\% of all false positives but only 5\% true positives. The area in which sources fulfill this condition is indicated by the dashed lines in Figure \ref{fig:distribution}.

\subsubsection{Point sources}

\begin{figure}[t]
\begin{center}
\includegraphics[width=1\columnwidth,trim=12 0 10 0,clip=true]{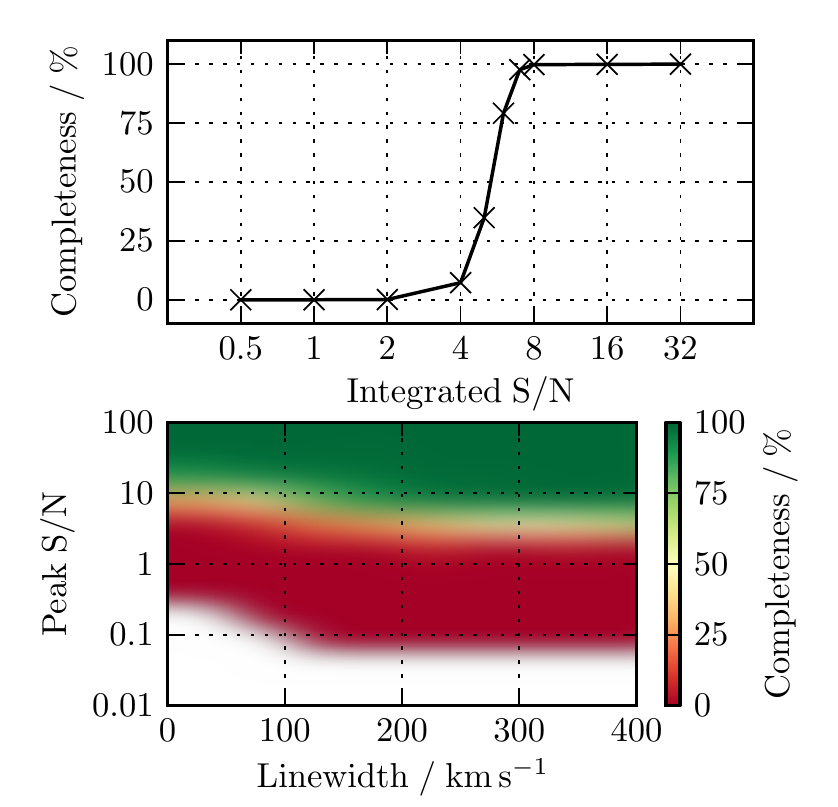}
\caption{Results from the simulation with spatially unresolved sources. The top panel shows the completeness as a function of the ISNR as probed in our simulations. The lower panel shows the completeness as a function of (logarithmic) peak signal-to-noise ratio and line width of the source. The white areas in the lower plot have not been tested. The reliability for this plot is 99\%.}
\label{fig:pointcutplot}
\end{center}
\end{figure}

We tested the completeness as a function of ISNR for both extended sources as well as point sources. To obtain realistic line profiles for the point sources, the extended models were summed up in each channel and convolved the resulting spectrum with the beam. The resulting point source model was then scaled to the desired ISNR.

The results of the run are summarized in Figure \ref{fig:pointcutplot}. Starting from ISNR 0.5, we increased the ISNR by a factor of two from bin to bin. Since the drop in completeness between ISNR 8 and 4 is rather sharp, we ran additional simulations in between those values.

The lower panel in Figure \ref{fig:pointcutplot} shows, that the source finder is indeed sensitive to the extended signal of the sources as we detect sources with a larger line width but lower peak signal-to-noise than the sources we do not detect for smaller line widths.

The reliability for these results is close to 100\%. We achieved this by applying the cut discussed in the previous section. Note that Figure \ref{fig:distribution} was made from the run with resolved sources.

\subsubsection{Extended sources}\label{sec:simres}

\begin{figure}[ht]
\begin{center}
\includegraphics[width=1\columnwidth,trim=12 0 10 0,clip=true]{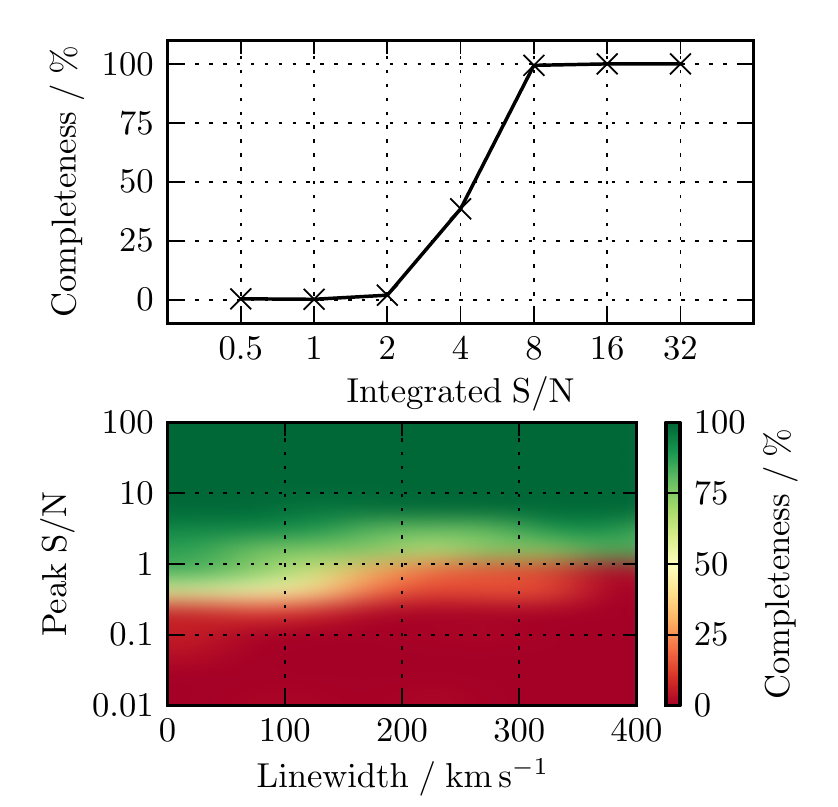}
\caption{Same as Figure \ref{fig:pointcutplot} for spatially extended sources. The reliability for this plot is 97\% and the fragmentation 3\%.}\label{fig:extcutplot}
\end{center}
\end{figure}

The second run was made with the extended galaxies which are clearly resolved by the simulated observations. Figure \ref{fig:extcutplot} shows the results of this run in a similar fashion to Figure \ref{fig:pointcutplot}. We again cut at $F_{R}$ and $F_{D} \leq 10\rm\,mJy\,km\,s^{-1}$ to reach a reliability of 97\%.

The boundary from 100\% completeness to 0\% is substantially smoother than in the case of the point sources. This behavior comes from the fact, that extended sources can be extended in the angular domain while at the same time being very narrow in the spectral domain, e.g. a galaxy seen face-on. This makes it substantially easier to detect galaxies with narrow line widths as long as they are extended in the angular domain. This is also evident from the lower panel in Figure \ref{fig:extcutplot}, where one can see, that narrow line width galaxies are detected to a lower peak signal-to-noise ratio than large line width galaxies.

For this simulation, we also encounter a phenomenon usually called fragmentation. We calculate it as the percentage of sources that have been detected two or more times. This can occur when a source with a very large line width is split into two detections. Furthermore, as mentioned in Section \ref{sec:scaleselect}, wavelet denoising schemes are generally prone to produce artifacts during the denoising process. Because of the simple way we determine whether a source is a true or a false detection, those artifacts can also cause multiple detections of the same source. We are therefore confident, that the fragmentation rate will decrease somewhat as the object identification process improves.

\section{Summary}

We have shown how 2D-1D wavelet denoising schemes can be used for source finding. Even with very simple post-processing of the denoised data, we set up an efficient source finding pipeline. Especially the robustness of the algorithm seems promising that it will work well on real data, which is certainly the next test to be passed.

Even though the splitting of the wavelet transformation in a 2D and 1D part avoids some of the difficulties that arise with anisotropic sources, it is far from perfect. A better denoising would be the usage of a full 3D curvelet transformation \citep{Candes:2007wf,2005SPIE.5914..351Y}. This transformation is however computationally much more difficult and demanding on the available hardware. We therefore think that our approach is a good compromise between sensitivity and computational complexity. But with more powerful hardware or more optimized algorithms, denoising by usage of the curvelet transform might become feasible, even for the large data sets we expect from the future radio telescopes.

Furthermore we like to stress that even though this algorithm was developed with H\,{\sc i} surveys in mind, it will in principle work for every kind of data that is similar to the data product of such a survey, e.g. other spectral line surveys.

\section*{Acknowledgments} 

\noindent
The authors thank the Deutsche Forschungsgemeinschaft (DFG) for support under grant number \mbox{KE757/7-2}.

\noindent
L. F. is member of the International Max Planck Research School (IMPRS) for Astronomy and Astrophysics at the Universities of Bonn and Cologne.

\noindent
Part of this work was done by L. F. as a summer student at the Netherlands Institute for Radio Astronomy (ASTRON) and he thanks his supervisors Tom Oosterloo, Gyula J\'ozsa and Paolo Serra for their support.

\noindent
This research has made use of NASA's Astrophysics Data System.

\bibliography{references}

\begin{thebibliography}{19}
\expandafter\ifx\csname natexlab\endcsname\relax\def\natexlab#1{#1}\fi

\bibitem[{{Atwood} {et~al.}(2009){Atwood}, {Abdo}, {Ackermann}, {Althouse},
  {Anderson}, {Axelsson}, {Baldini}, {Ballet}, {Band}, {Barbiellini}, \&
  et~al.}]{2009ApJ...697.1071A}
{Atwood}, W.~B., {Abdo}, A.~A., {Ackermann}, M., {et~al.} 2009, \apj, 697, 1071

\bibitem[{{Barnes} {et~al.}(2001){Barnes}, {Staveley-Smith}, {de Blok},
  {Oosterloo}, {Stewart}, {Wright}, {Banks}, {Bhathal}, {Boyce}, {Calabretta},
  {Disney}, {Drinkwater}, {Ekers}, {Freeman}, {Gibson}, {Green}, {Haynes}, {te
  Lintel Hekkert}, {Henning}, {Jerjen}, {Juraszek}, {Kesteven}, {Kilborn},
  {Knezek}, {Koribalski}, {Kraan-Korteweg}, {Malin}, {Marquarding}, {Minchin},
  {Mould}, {Price}, {Putman}, {Ryder}, {Sadler}, {Schr{\"o}der}, {Stootman},
  {Webster}, {Wilson}, \& {Ye}}]{2001MNRAS.322..486B}
{Barnes}, D.~G., {Staveley-Smith}, L., {de Blok}, W.~J.~G., {et~al.} 2001,
  \mnras, 322, 486

\bibitem[{Candes {et~al.}(2007)Candes, Demanet, \& Donoho}]{Candes:2007wf}
Candes, E., Demanet, L., \& Donoho, D. 2007, Multiscale modeling and {\ldots}

\bibitem[{{Giovanelli} {et~al.}(2005){Giovanelli}, {Haynes}, {Kent},
  {Perillat}, {Saintonge}, {Brosch}, {Catinella}, {Hoffman}, {Stierwalt},
  {Spekkens}, {Lerner}, {Masters}, {Momjian}, {Rosenberg}, {Springob},
  {Boselli}, {Charmandaris}, {Darling}, {Davies}, {Garcia Lambas}, {Gavazzi},
  {Giovanardi}, {Hardy}, {Hunt}, {Iovino}, {Karachentsev}, {Karachentseva},
  {Koopmann}, {Marinoni}, {Minchin}, {Muller}, {Putman}, {Pantoja}, {Salzer},
  {Scodeggio}, {Skillman}, {Solanes}, {Valotto}, {van Driel}, \& {van
  Zee}}]{2005AJ....130.2598G}
{Giovanelli}, R., {Haynes}, M.~P., {Kent}, B.~R., {et~al.} 2005, \aj, 130, 2598

\bibitem[{{Holschneider} {et~al.}(1989){Holschneider}, {Kronland-Martinet},
  {Morlet}, \& {Tchamitchian}}]{1989wtfm.conf..286H}
{Holschneider}, M., {Kronland-Martinet}, R., {Morlet}, J., \& {Tchamitchian},
  P. 1989, in Wavelets. Time-Frequency Methods and Phase Space, ed.
  {J.-M.~Combes, A.~Grossmann, \& P.~Tchamitchian}, 286--+

\bibitem[{{Johnston} {et~al.}(2008){Johnston}, {Taylor}, {Bailes}, {Bartel},
  {Baugh}, {Bietenholz}, {Blake}, {Braun}, {Brown}, {Chatterjee}, {Darling},
  {Deller}, {Dodson}, {Edwards}, {Ekers}, {Ellingsen}, {Feain}, {Gaensler},
  {Haverkorn}, {Hobbs}, {Hopkins}, {Jackson}, {James}, {Joncas}, {Kaspi},
  {Kilborn}, {Koribalski}, {Kothes}, {Landecker}, {Lenc}, {Lovell}, {Macquart},
  {Manchester}, {Matthews}, {McClure-Griffiths}, {Norris}, {Pen}, {Phillips},
  {Power}, {Protheroe}, {Sadler}, {Schmidt}, {Stairs}, {Staveley-Smith},
  {Stil}, {Tingay}, {Tzioumis}, {Walker}, {Wall}, \&
  {Wolleben}}]{2008ExA....22..151J}
{Johnston}, S., {Taylor}, R., {Bailes}, M., {et~al.} 2008, Experimental
  Astronomy, 22, 151

\bibitem[{{Kerp} {et~al.}(2011){Kerp}, {Winkel}, {Ben Bekhti}, {Fl{\"o}er}, \&
  {Kalberla}}]{2011AN....332..637K}
{Kerp}, J., {Winkel}, B., {Ben Bekhti}, N., {Fl{\"o}er}, L., \& {Kalberla},
  P.~M.~W. 2011, Astronomische Nachrichten, 332, 637

\bibitem[{{Koribalski} \& {Staveley-Smith}(2009)}]{wallabyprop}
{Koribalski}, B.~S. \& {Staveley-Smith}, L. 2009, ASKAP Survey Science Proposal

\bibitem[{{Koribalski} {et~al.}(2004){Koribalski}, {Staveley-Smith}, {Kilborn},
  {Ryder}, {Kraan-Korteweg}, {Ryan-Weber}, {Ekers}, {Jerjen}, {Henning},
  {Putman}, {Zwaan}, {de Blok}, {Calabretta}, {Disney}, {Minchin}, {Bhathal},
  {Boyce}, {Drinkwater}, {Freeman}, {Gibson}, {Green}, {Haynes}, {Juraszek},
  {Kesteven}, {Knezek}, {Mader}, {Marquarding}, {Meyer}, {Mould}, {Oosterloo},
  {O'Brien}, {Price}, {Sadler}, {Schr{\"o}der}, {Stewart}, {Stootman}, {Waugh},
  {Warren}, {Webster}, \& {Wright}}]{2004AJ....128...16K}
{Koribalski}, B.~S., {Staveley-Smith}, L., {Kilborn}, V.~A., {et~al.} 2004,
  \aj, 128, 16

\bibitem[{{Murtagh} {et~al.}(1995){Murtagh}, {Starck}, \&
  {Bijaoui}}]{1995A&AS..112..179M}
{Murtagh}, F., {Starck}, J.-L., \& {Bijaoui}, A. 1995, \aaps, 112, 179

\bibitem[{Nyquist(1928)}]{nyquist}
Nyquist, H. 1928, Trans. AIEE, 47, 617

\bibitem[{{Oosterloo} {et~al.}(2009){Oosterloo}, {Verheijen}, {van Cappellen},
  {Bakker}, {Heald}, \& {Ivashina}}]{2009wska.confE..70O}
{Oosterloo}, T., {Verheijen}, M.~A.~W., {van Cappellen}, W., {et~al.} 2009, in
  Proceedings of Wide Field Astronomy \& Technology for the Square Kilometre
  Array (SKADS 2009). 4-6 November 2009. Chateau de Limelette, Belgium.

\bibitem[{Serra(1982)}]{Serra:1982fk}
Serra, J.~P. 1982, Image analysis and mathematical morphology (Academic Press)

\bibitem[{Starck \& Bobin(2010)}]{5299269}
Starck, J.-L. \& Bobin, J. 2010, Proceedings of the IEEE, 98, 1021

\bibitem[{Starck {et~al.}(2007)Starck, Fadili, \& Murtagh}]{4060954}
Starck, J.-L., Fadili, J., \& Murtagh, F. 2007, Image Processing, IEEE
  Transactions on, 16, 297

\bibitem[{{Starck} {et~al.}(2009){Starck}, {Fadili}, {Digel}, {Zhang}, \&
  {Chiang}}]{2009A&A...504..641S}
{Starck}, J.-L., {Fadili}, J.~M., {Digel}, S., {Zhang}, B., \& {Chiang}, J.
  2009, \aap, 504, 641

\bibitem[{Starck {et~al.}(2010)Starck, Murtagh, \& Fadili}]{Starck:2010uf}
Starck, J.-L., Murtagh, F., \& Fadili, J.~M. 2010, {Sparse Image and Signal
  Processing: Wavelets, Curvelets, Morphological Diversity} (Cambridge
  University Press)

\bibitem[{{Winkel} {et~al.}(2010){Winkel}, {Kalberla}, {Kerp}, \&
  {Fl{\"o}er}}]{2010ApJS..188..488W}
{Winkel}, B., {Kalberla}, P.~M.~W., {Kerp}, J., \& {Fl{\"o}er}, L. 2010, \apjs,
  188, 488

\bibitem[{Ying {et~al.}(2005)Ying, Demanet, \& Candes}]{2005SPIE.5914..351Y}
Ying, L., Demanet, L., \& Candes, E. 2005, Wavelets XI. Edited by Papadakis,
  5914, 351

\end{thebibliography}


\end{document}